\documentclass[useAMS,usenatbib,usegraphicx]{mn2e}

\def\arcsec{\hbox{$^{\prime\prime}$}}

\title[Diamonds and PAHs in the Circumstellar Environment of the Herbig Ae/Be Star Elias~1]{Diamonds and PAHs in the Circumstellar
Environment of the Herbig Ae/Be Star Elias~1}

\author[R. Topalovic, J. Russell, J. McCombie, T.H. Kerr and P.J. Sarre]{R. Topalovic$^{1}$, J. Russell$^{1}$, J. McCombie$^{1}$, T.H. Kerr$^{2}$ and P. J. Sarre$^{1}$\\
$^{1}$School of Chemistry, The University of Nottingham, University Park, Nottingham NG7 2RD, U.K.\\
$^{2}$United Kingdom Infrared Telescope, Joint Astronomy Centre, 660 N. A'ohoku Place, University Park, Hilo, Hawaii
96720, U.S.A.}

\begin{document}


\pagerange{\pageref{firstpage}--\pageref{lastpage}} \pubyear{2006}

\maketitle

\label{firstpage}

\begin{abstract}
We report long-slit spectroscopic observations of the Herbig Ae/Be
star Elias~1 in the 3.2 - 3.6$\,\umu$m region covering the C-H
stretch emission features of hydrogen-terminated diamonds and PAHs.
The data were recorded at UKIRT using UIST and yield information on
the profiles and intensities of the bands as a function of offset
along the N-S and E-W axes centred on the close binary. The diamond
and nearby IR continuum emission arises from a symmetrical inner
core region ($\leq$\,0.34$^{\prime\prime}$ or 48 AU).  The
3.3$\,\umu$m PAH emission is extended along the E-W axis up to
\emph{c.} 100 AU each side of the star.  This result supports a
suggestion of \citet{haa97} of an E-W oriented bipolar nebula in
Elias 1.

\end{abstract}

\begin{keywords}
stars: circumstellar matter --stars: binaries --stars: individual:
V892 Tau
\end{keywords}

\section{Introduction}
The morphology, evolution and chemical composition of circumstellar
disks and halos of pre-main-sequence intermediate-mass Herbig Ae/Be
stars is of key importance in the study of star and planet formation
\citep{wat98}.  Emission features of silicates and PAHs are commonly
though not always observed \citep{ack04} and in a few cases emission
from surface-hydrogen-terminated diamonds is also detected. This
raises interesting and challenging questions as to how the diamonds
form, why their emission is seen in only a few stars, and whether
there is a link between circumstellar diamonds and the presence of
nanodiamonds in meteorites.  The existence of organic PAHs in the
nebular environment also provides astrobiological interest. In this
paper we describe near-IR long-slit observations of one of the rare
diamond C-H emitting objects, Elias~1. We focus on the spectroscopy
and spatial distribution of diamond (C-H), PAH and continuum
emission from the circumstellar environment, and make comparison
with the related object HD~97048.

The pre-main-sequence star Elias~1 (V892 Tau, Elias~3-1) is a close
binary with a separation of $\sim$50 mas at PA $\sim$50$^\circ$
\citep{smi05}. Its spectral type is uncertain ranging through A0
\citep{eli78}, A6 with $A$$_V$~=~3.9 mag \citep{zin94}, B9 with
$A$$_V$~=~8.85 mag \citep{str94},
 with an $A_V$ value of 10.5 also reported \citep{tei99}. Elias~1
 has a very flat, possibly rising, SED in
the near-IR to far-IR which is markedly different from that of
HD~97048 \citep{hil92}, the two stars being classified as group II
and I, respectively.  The SED for Elias~1 is taken by \citet{smi05}
to suggest the possible existence of an envelope in addition to a
disk, or a flared disk. Other Elias~1 characteristics include X-ray
\citep{zin94,gia04,ham05,ste06} and radio \citep{ski93} emission.

An emission feature in the \emph{c.} 3.5$\,\umu$m region for Elias~1
was first reported in 1982 by \citet{all82} and was subsequently
shown to have two components near 3.41 and 3.52$\,\umu$m with an
additional PAH feature at 3.29$\,\umu$m \citep{whi83}.  The only
previously detected example was HD~97048 \citep{bla80}.  Subsequent
spectroscopic observations of Elias~1 have been made
\citep{whi84,tok91,geb97}, including ISO (SWS) data \citep{vank02}.
The features at 3.43 and 3.53$\,\umu$m remained unassigned until
\citet{gui99} found a convincing correspondence between the
astrophysical bands of Elias~1 and HD~97048 and laboratory
absorption spectra of the C-H stretching modes on
hydrogen-terminated diamond nanocrystal films at $\sim$~1000~K
\citep{cha95}. ISO spectra of Elias~1 covering the range 2.5~-~13.5
$\,\umu$m \citep{vank02} contain absorption features due to water
and CO$_2$ ice, and emission from PAHs, diamonds and silicates, the
last of these being absent from the spectrum of HD~97048. In this
paper we have chosen not to use the word `nanodiamond' in discussing
the stellar emission features as their size is probably at least
50~nm \citep{che02,she02,jon04} in contrast to meteoritic
nanodiamonds which are typically only 2-3~nm across \citep{lew87}.
We also
 remark that the diamonds observed are exclusively those that are
 H-terminated at the surface; absence of hydrogen or its substitution
 by other chemical groups on the surface in general give spectra outside the region
 studied. The only other Herbig Ae/Be objects for
 which diamond emission is reported are BD$+$40$^\circ$4124 \citep{ack04}
 and MWC 297 \citep{ter01,ack04}; emission features attributed to diamond emission
 are also seen in the post-AGB star HR~4049 \citep{geb89} but are
 relatively weak.

\section{The Circumstellar Environment of Elias~1}

Unresolved at mm wavelengths \citep{dif97,hen98}, observations of
the spatial distribution of the circumstellar emission of Elias~1
comprise a contour map of the inner 0.2$^{\prime\prime}$ $\times$
0.2$^{\prime\prime}$ region which exhibits asymmetry along the
binary PA of 50$^{\circ}$ \citep{smi05}, and speckle interferograms
at K, L (0$^{\circ}$ and 90$^{\circ}$) and L at 45$^{\circ}$
\citep{kat91} and at J, H, K and L$^\prime$ \citep{haa97}.
\citet{kat91} interpreted their data in terms of a core with a
wavelength-dependent diameter of $\sim$ 6-14 AU, and a highly
flattened halo elongated in the east-west direction with radius
$\sim$ 30-100 AU in the form of a nearly edge-on circumstellar disk.
Based on observations covering a wider range of wavelengths and at
higher resolution, \citet{haa97} proposed that the east-west
elongation is due to scattered light from bipolar (nebular) lobes.
From the PA 90$^{\circ}$ and 0$^{\circ}$ data it was concluded that
the components observed at J and H comprise a well-resolved narrow
blue structure elongated east-west with a FWHM of \emph{c.}
1$^{\prime\prime}$ and a second that is marginally resolved (FWHM
$\leq$\,0.2$^{\prime\prime}$), and that there is marginally resolved
red structure at K and L$^{\prime}$ which is more circular and
symmetric. From a consideration of all the available data and a
number of possible interpretations, \citet{haa97} proposed the
existence of a bipolar nebula along the E-W axis as the origin of
the E-W extension. They also discussed the existence of a possible
stellar companion, noting that their K and L$^\prime$ data could be
fitted by a close binary with separation oriented at PA 45$^{\circ}$
or 135$^{\circ}$ with separation $<$\,0.1$^{\prime\prime}$, together
with some extended structure. Recent diffraction-limited speckle
interferometric observations confirm that Elias~1 is indeed a close
binary system with these characteristics, the separation being
0.05$^{\prime\prime}$ with PA 50$^{\circ}$ \citep{smi05}.\\

\section[]{Observations}

Observations of the Herbig Ae/Be star Elias 1 were carried out
between 20th July - 11th September 2004 at the 3.8~m United Kingdom
Infrared Telescope (UKIRT). The 1 - 5 $\mu$m imager-spectrometer
(UIST) was used for the long-slit spectroscopic observations. The
slit dimensions were 0.48$^{\prime\prime}$ x 120$^{\prime\prime}$
and the resolving power = 700. The short-L grism was used to obtain
spectra in the wavelength region 2.905 - 3.638$\,\umu$m in
combination with an InSb array of 1024 x 1024 pixels, giving a
spatial resolution of 0.12$^{\prime\prime}$ pix$^{-1}$. The slit was
centred on the star and orientated at position angles 0$^\circ$ (NS)
and 90$^\circ$ (WE). The observational details are listed in table
1. The sky background emission was subtracted from the standard and
object frames and the object spectra were divided by the standard.
The point spread function (PSF) of the observations is very
consistent. A further exposure along E-W was made on 27 July but due
to poor seeing the PSF was \emph{c.} 0.7$^{\prime\prime}$ and was
not included in this analysis.

The data were partially reduced by the ORAC-DR (automated reduction)
pipeline for UIST and further reductions were carried out with the
FIGARO reduction package. The spectra were analysed using the IRAF
reduction package; the continuum was fitted and subtracted from the
spectra and Lorentzians were used to fit the emission bands.

\begin{table}
\begin{minipage}{84mm}
\caption{Observing log} \label{tab:ol}
\begin{tabular}{lcllll}
\hline
        &        Date  & slit        & exp       & standard & PSF\\
      &       (2004)  &     & (sec)     & & FWHM($^{\prime\prime}$)
\\
\hline
Elias~1 &       Jul 20  & N-S      & 80    &  BS 1203 & 0.53\\
Elias~1 &       Jul 28  & N-S      & 128    &  BS 1203 & 0.52\\
Elias~1 &       Sep 11  & N-S      & 80    &  BS 1203 & 0.55\\
Elias~1 &       Sep 11  & E-W      & 80    &  BS 1203 & 0.52\\
\hline
\end{tabular}

\end{minipage}
\end{table}

\section{Infrared spectroscopic features and spatial distributions}

\subsection{H-terminated diamond and PAH spectra}

In common with similar long-slit studies of Red Rectangle PAH
features (Song et al. 2003), one of the objectives was to
investigate any evolution in profile shape, central wavelength etc.
of the PAH C-H stretch and (particularly) the C-H diamond features
with increase in offset from the star. The summed 3.0 - 3.6$\,\umu$m
spectra for Elias~1 along the N-S axis for the inner
$\pm\,0.3\arcsec$ relative to the star are shown in Fig 1 and are
similar to the UKIRT data of \citep{geb97} although the Pf$\delta$
line of hydrogen which sits atop the 3.3$\,\umu$m PAH band is much
weaker in our spectra. The Pf$\delta$ flux in our 2004 data is
\emph{c.} 0.2 x 10$^{-13}$ Wm$^{-2}$$\umu$m$^{-1}$ compared with
\emph{c.} 1.0 x 10$^{-13}$ Wm$^{-2}$$\umu$m$^{-1}$ \citep{geb97}.
The assignments of the C-H diamond features have been described
elsewhere \citep{gui99,vank02} and are not reproduced here. The best
match between experimental absorption spectra and the astrophysical
data is achieved for a temperature of \emph{c.} 1000~K
\citep{gui99,vank02}. Laboratory data have revealed a strong
dependence of the spectrum on particle size, indicating that the
diamonds in Elias~1 (and HD~97048) are probably at least 50~nm in
diameter and are likely formed in a CVD-type process
\citep{che02,she02,jon04}.

\begin{figure}
\includegraphics[scale=0.34]{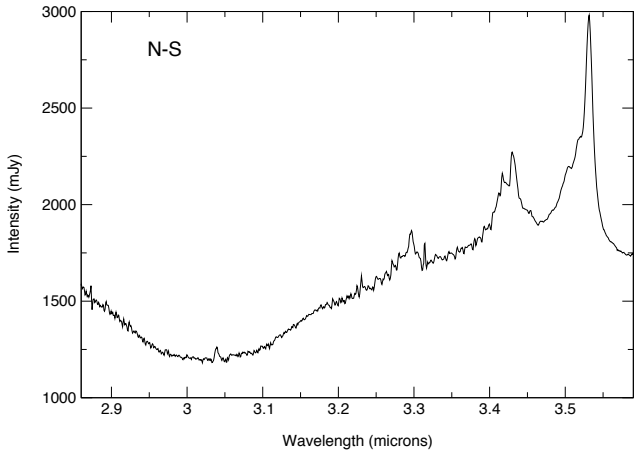}
\caption{Co-added spectra of Elias~1 in the 2.9~-~3.6$\,\umu$m
region recorded with slit alignment N-S .  The data were summed over
the inner $\pm\,0.3\arcsec$.  The data for E-W are the same within
the signal-to-noise ratio.} \label{fig1}
\end{figure}

As described in the following sub-section, the diamond distribution
in Elias~1 is confined to a region quite close to the star within
$\leq$\,48 AU (FWHM, Gaussian fitted). By taking data extractions
`on-star' at the point of maximum emission strength
 and secondly at the greatest possible offset for which a reasonable signal-to-noise
 is still maintained, we searched for variation in the diamond features but did
 not find evidence for this to within 0.15\% (in $\lambda$), 20\% (in FWHM), and 10\% (in relative intensity).
 The PSF of \emph{c.} 0.5$^{\prime\prime}$ is larger than desirable
relative to the diamond
 emission spatial extent of FWHM $\leq$\,0.34$^{\prime\prime}$.  However, changes in the relative band strengths of the components,
 in peak wavelength or in profile shape that might occur due to change in temperature with offset,
 diamond size effects on the spectra, or in the relative emission contributions of
 the different C-H supporting diamond facets, would be discernible.  This
 result points to a rather specific set of formation
 conditions and/or surface chemistry, possibly not unrelated to the rarity of the
 appearance of the features.

Of the main PAH emission bands, only the relatively weak
3.3$\,\umu$m
 feature occurs in the spectral range recorded and is
subject to contamination by telluric methane and water lines, as
well as from the Pf$\delta$ emission line of the star.  As for the
diamond emission bands there was no measurable change in the feature
profile with offset but a 20\% change in width would have been
detected.

\subsection{Spatial distributions of PAH, diamond and continuum emission}

Slit alignments along the N-S and the E-W axes were selected in
order to probe the spatial distribution of the PAH (3.3$\,\umu$m),
diamond and continuum emission relative to the star.  Although
adaptive optics and larger mirror size (and speckle interferometric
observations) offer a lower effective PSF, high signal-to-noise data
achieved with moderate exposure times and at high altitude on Mauna
Kea provide an attractive option with relatively straightforward
data reduction procedures. Determination of the PSF was achieved
using the standard (point source) star BS 1203 and gave a stable
value of 0.53(2)$^{\prime\prime}$ (see table 1.)

The N-S and E-W spatial distributions as recorded for the PAH
(3.3$\,\umu$m, continuum-subtracted), diamond (3.53$\,\umu$m,
continuum-subtracted) and continuum (3.60$\,\umu$m) are given in
figs 2 and 3, respectively. The set of diamond features near
3.43$\,\umu$m follow the 3.53$\,\umu$m distribution within
signal-to-noise contraints, the measured spatial FWHM for
3.53$\,\umu$m being 0.590$^{\prime\prime}$ (NS) and
0.603$^{\prime\prime}$ (EW), and 0.603$^{\prime\prime}$ (NS) and
0.597$^{\prime\prime}$ (EW) for the 3.43$\,\umu$m group. Figs 4 and
5 show the results when a Richardson-Lucy (RL) deconvolution
algorithm was employed to deconvolve the profiles. For the N-S slit
position (PA~=~0$^\circ$) in fig 4 the profiles for PAH, diamond and
continuum overlap very well for the inner $\pm\,~0.2$\arcsec, with
the latter two overlapping for all offsets. However, the PAH
distribution extends weakly about three times further, with slightly
greater strength in the N direction. For the E-W slit position
(PA~=~90$^\circ$) shown in figure 5 the diamond and continuum
distributions are the same within the signal-to-noise. In contrast,
the PAH distribution is very broad and skewed slightly towards W.

The deconvolved continuum-subtracted 3.53$\,\umu$m Elias~1 diamond
feature shows a symmetrical distribution N, S, E and W of the
central star. This is consistent with the circular symmetric
component described by \citet{haa97} for K and L$^{\prime}$ who
noted that the latter filter at 3.69~$\pm$\,~0.315~$\,\umu$m
includes the strong 3.42 and 3.53$\,\umu$m features observed in our
study and that are now known to be due to diamond emission. The
distribution for diamond emission has a FWHM of 0.34$\arcsec$ giving
a Gaussian FWHM diameter of $\leq$\,48 AU where a distance to
Elias~1 of 140 pc \citep{eli78} was adopted. The continuum profile
that is generally taken to arise from thermal emission by larger
grains is very similar.

The PAH distribution is extended relative to diamond and continuum
emission along both slit axes but more so for E-W than N-S.  For E-W
a signal is discernible to $\pm\,$0.7$\arcsec$ or $\pm\,$100 AU W
and E relative to the star. This long-slit spectroscopic result
agrees well with the findings of \citet{haa97} who found a narrow
E-W elongated blue halo in J and H with FWHM of about 1$\arcsec$,
the N-S extent only being about 0.2$\arcsec$.  We conclude that the
new data provide strong evidence for a PAH-containing bipolar nebula
oriented close to E-W but, given the asymmetry in the PAH
distribution along both E-W and N-S (see figs 4 and 5), tilted
slightly with respect to the observer.

\begin{figure}
\includegraphics[scale=0.34]{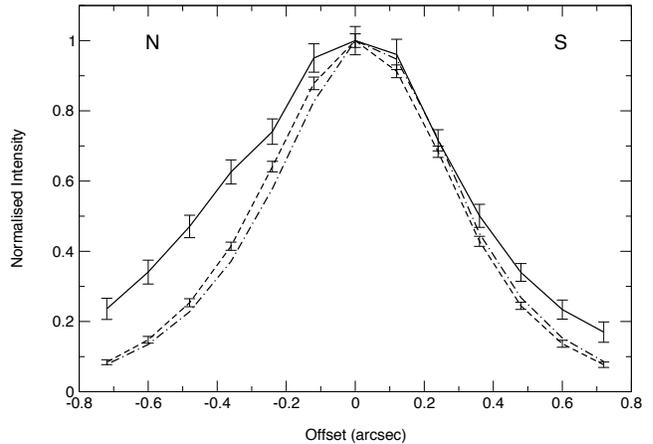}
\caption{Normalised emission intensities along the N-S axis for the
PAH (3.3$\,\umu$m) - solid line, diamond (3.53$\,\umu$m) - dashed
line, and continuum (3.60$\,\umu$m) - dot-dashed line.} \label{fig2}
\end{figure}

\begin{figure}
\includegraphics[scale=0.34]{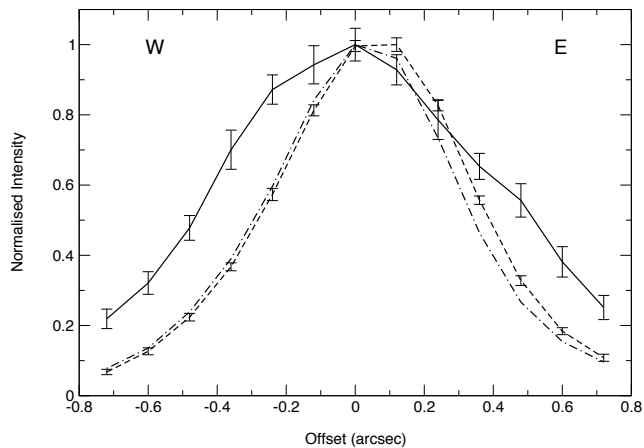}
\caption{Normalised emission intensities along the E-W axis for the
PAH (3.3$\,\umu$m) - solid line, diamond (3.53$\,\umu$m) - dashed
line, and continuum (3.60$\,\umu$m) - dot-dashed line.} \label{fig3}
\end{figure}

\section{Comparison of Elias~1 with HD~97048}

It is of interest to compare the results for Elias~1 with those for
HD~97048.  There is a clear similarity in that they both exhibit
PAH, diamond and IR continuum emission, but they have very different
SEDs and fall in different classifications being in Group II with a
large IR excess (Elias~1) and in Group I (HD~97048) with a
circumstellar disk seen pole-on \citep{hil92,hab04}.

 Based on
speckle observations, \citet{roc86} reported that the 3.53$\,\umu$m
diamond emission of HD~97048 is not substantially extended, arising
within 0.1$^{\prime\prime}$ of the star. \citet{hab04} more recently
described an adaptive optics high angular resolution (\emph{c.}
0.1$^{\prime\prime}$) long-slit spectral study of the spatial
distribution of diamond and continuum emission for HD~97048. It was
deduced that the FWHM sizes are consistent with predictions for a
circumstellar disk oriented pole-on, with the diamond emission
originating in an inner region.  The diamond FWHM emission for
HD~97048 is reported by \citet{hab06} to be 41 AU and that of the
continuum 32 AU, which compare with our Elias~1 results of
$\leq$\,0.34$^{\prime\prime}$ (48 AU) for both diamond and the
continuum.

 \citet{hab06} have reported spatially resolved 3.3$\,\umu$m
 PAH emission for HD~97048 which is more extended than the diamond and continuum emission
 and hence similar to our results for Elias~1. The PAH distribution data for HD~97048 is incomplete due to
a noisy section near 3.3$\,\umu$m falling in the +0.0 -
0.2$^{\prime\prime}$ region and a value for the FWHM was not quoted.
However, from the negative offset data (figure 3 of the paper) and
assuming a symmetrical spatial profile, a FWHM of $\sim$\,60 AU can
be estimated. An extended 3.3$\,\umu$m distribution for HD~97048 is
also consistent with a comparison between the ISO-SWS
(20$^{\prime\prime}$ x 33$^{\prime\prime}$) spectrum and the spectra
of Habart et al. (2004) within 1$^{\prime\prime}$ of the star which
showed the PAH 3.3$\,\umu$m band to be stronger relative to the
diamond bands in the ISO spectrum. Resolved PAH profiles in other
Herbig Ae/Be stars have been found and successfully modelled by
\citet{hab04a}.  In general the 3.3$\,\umu$m PAH feature tends to be
more confined spatially than the UIR bands at 6.2, 7.7, 8.6 and
11.2$\,\umu$m. This is so for HD 97048 which extends \emph{c.} 5 -
10$\arcsec$ in the mid-IR \citep{sie00} and from limited nulling
interferometric observations is reported to be the case for Elias~1
\citep{liu05}.

\begin{figure}
\includegraphics[scale=0.34]{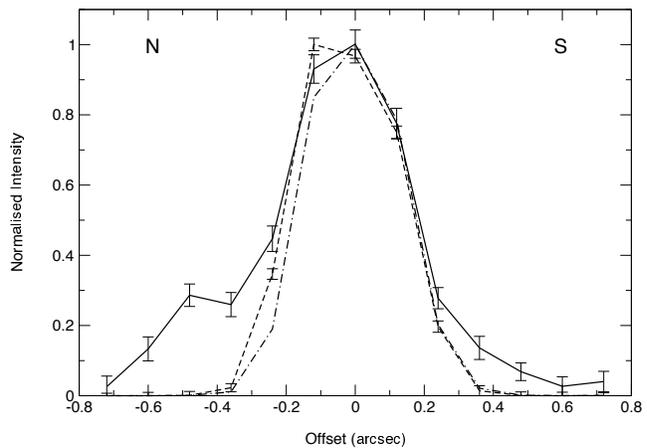}
\caption{Normalised emission intensities along
the N-S axis for the PAH (3.3$\,\umu$m) - solid line, diamond
(3.53$\,\umu$m) - dashed line, and continuum (3.60$\,\umu$m) -
dot-dashed line, where the data for different nights (see table 1.)
are co-added and a RL deconvolution applied.} \label{fig4}
\end{figure}

\begin{figure}
\includegraphics[scale=0.34]{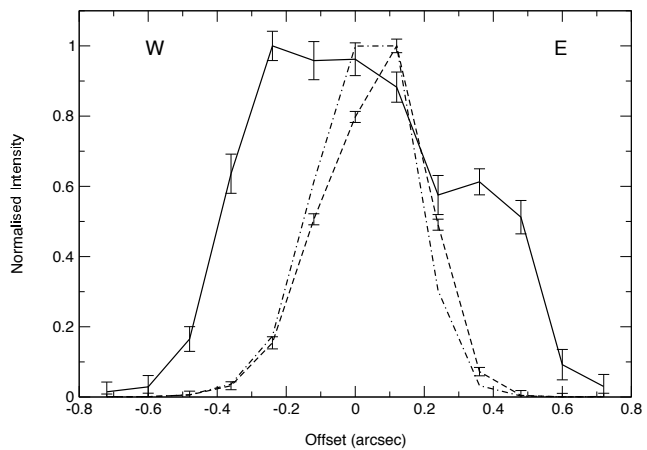}
\caption{Normalised emission intensities along the E-W axis for the
PAH (3.3$\,\umu$m) - solid line, diamond (3.53$\,\umu$m) - dashed
line, and continuum (3.60$\,\umu$m) - dot-dashed line, where a
 RL deconvolution has been applied.} \label{fig5}
\end{figure}

\section{Summary}

We have conducted a long-slit study of Elias~1 in the 3.2 -
3.6$\,\umu$m region which has revealed a symmetrical distribution
for the continuum (3.6$\,\umu$m) and diamond C-H emission features,
and more extended PAH 3.3$\,\umu$m emission which is oriented along
the E-W axis.  Given that for many objects the 3.3$\,\umu$m PAH
emission is significantly less extended than for the UIR bands at
longer wavelengths, it will be of interest to determine the mid-IR
PAH emission distribution using long-slit spectroscopy in order to
probe the size and geometry of the bipolar nebula.

\section{Acknowledgments}

We thank the UK Panel for the Allocation of Telescope Time for the
award of observing time on UKIRT, PPARC for a studentship to RT and
EPSRC for a studentship to JR.

\label{lastpage}

\end{document}